\documentclass{article}
\usepackage[english]{babel}
\usepackage[utf8]{inputenc}
\usepackage{capt-of}
\usepackage{graphicx}
\usepackage{subfig}
\usepackage{authblk}
\usepackage{wrapfig}
\usepackage{amsmath,amssymb}
\usepackage[%
    colorlinks=true,
    pdfborder={0 0 0},
    linkcolor=red
]{hyperref}
\usepackage[
backend=biber,
style=ieee,
sorting=none
]{biblatex}

\title{Invariant mass reconstruction of heavy gauge bosons decaying to $\tau$ leptons using machine learning techniques}

\author[1,2]{Vinaya Krishnan MB\footnote{vkrishnan@iopb.res.in}}\author[1,2]{Aruna Kumar Nayak\footnote{nayak@iopb.res.in}}\author[3]{Asrith Krishna Radhakrishnan\footnote{asrithkr@gmail.com}}
\affil[1]{Institute of Physics, Bhubaneswar, India }
\affil[2]{Homi Bhabha National Institute, Training School Complex, Anushakti Nagar, Mumbai 400 085, India}
\affil[3]{Indian Institute of Science Education and Research Mohali, India}

\date{}

\begin{document}
\maketitle
\begin{abstract} 
\noindent  Many analyses are performed by the LHC experiments to search for heavy gauge bosons, which appear in several new physics models. The invariant mass reconstruction of heavy gauge bosons is difficult when they decay to $\tau$ leptons due to missing neutrinos in the final state. Machine learning techniques are widely utilized in experimental high-energy physics, in particular in analyzing the large amount of data produced at the LHC. In this paper, we study machine learning techniques such as supervised and unsupervised neural network algorithms to reconstruct the invariant mass of $Z^{\prime}~\rightarrow~\tau\tau$ and $W^{\prime}~\rightarrow~\tau\nu$ decays, which can improve the sensitivity of these searches. \end{abstract}

\section{Introduction}
After the discovery of a Standard Model (SM) like Higgs boson by the ATLAS and CMS experiments at CERN-LHC \cite{atlas,atlas1}, a major focus of the current physics program at LHC is to search for beyond Standard Model (BSM) signatures. Many analyses are performed at LHC to search for the production of heavy gauge bosons, $Z^{\prime}$ and $W^{\prime}$, which are predicted by several BSM models. In certain BSM scenarios, the $Z^{\prime}$ and $W^{\prime}$ bosons can preferentially decay to third-generation fermions, which motivates the search for $Z^{\prime}~\rightarrow~\tau\tau$ and $W^{\prime}~\rightarrow~\tau\nu$ decays. Both ATLAS and CMS experiments have performed searches for $Z^{\prime}$ and $W^{\prime}$ bosons in the final states with $\tau$ leptons \cite{exo1,exodd,exo3}. 

The $\tau$ lepton, being the heaviest among the three leptons, decays after a short time to electron, muon, or hadrons, accompanied by neutrinos. The charged leptons, charged hadrons, and neutral hadrons (mostly $\pi^{0}$s, which further decay to a pair of photons) can be observed in the detector and are referred to as visible decay products. Neutrinos, being very weakly interacting particles, escape the detector undetected. However, at the ATLAS and CMS experiment, the sum of the transverse momentum of neutrinos is indirectly inferred from the momentum imbalance in the transverse plane. So, the missing transverse momentum ($\vec{p}_{T}^{miss}$) is defined as the negative vector sum of all visible particles in the transverse plane. But, the z-component of the neutrino momentum can not be measured since the momentum of the colliding partons is not known. Therefore, it is difficult to reconstruct the invariant mass of the $Z^{\prime}~\rightarrow~\tau\tau$ and 
$W^{\prime}~\rightarrow~\ell\nu$ ($\ell~={\rm e},~\mu,~{\rm or}~\tau$) decays. The ATLAS and CMS experiments have used visible di-$\tau$ mass ($m^{vis}_{\tau\tau}$, reconstructed from the visible component of the $\tau$ momentum) or $m(\tau_1, \tau_2, \vec{p}_{T}^{miss})$ \cite{exodd} in searches for $Z^{\prime}~\rightarrow~\tau\tau$ and $m_{T}(\ell, \vec{p}_{T}^{miss})$ in searches for $W^{\prime}~\rightarrow~\ell\nu$, respectively \cite{exo1,exodd,exo3}. 

Recently, machine learning (ML) techniques have been widely applied in HEP data analyses, especially at LHC, providing remarkable improvements in particle identification, jet classification, event classification, energy regression, etc. In this article, we study the application of machine learning in the context of reconstructing the full invariant mass of the $Z^{\prime}~\rightarrow~\tau\tau$ and $W^{\prime}~\rightarrow~\tau\nu$ decays, which can help in improving the sensitivity of these searches at LHC due to better separation of signal from the SM backgrounds. 
We study the reconstruction of $Z^{\prime}~\rightarrow~\tau\tau$ invariant mass using an artificial neural network (NN), which is implemented with the Python deep learning library "Keras"\cite{keras}. We compare the performance of our results to that of $m^{vis}_{\tau\tau}$ and the invariant mass computed using SVfit\cite{SVfit}, which is based on the likelihood method and used in CMS experiment to compute the invariant mass of  $Z~\rightarrow~\tau\tau$ and $H~\rightarrow~\tau\tau$ decays. 
To reconstruct the invariant mass of $W^{\prime}~\rightarrow~\tau\nu$ we study a generative adversarial network. The method can also be applicable for $W^{\prime}~\rightarrow~{\rm e}\nu$ and $W^{\prime}~\rightarrow~\mu\nu$ searches. 

The rest of the paper is structured as follows. The details of BSM and SM simulated samples and the control regions are described in section 2. Sections 3 and 4 provide the details about the machine learning techniques used for the reconstruction of the invariant mass of $Z'\rightarrow\tau\tau$ and $W'\rightarrow\tau\nu_{\tau}$ final states. The results are presented in terms of improvement in mass resolution. No effort is made to provide any signal significance since a full search analysis is beyond the scope of this paper.

\section{Event generation}
 Monte Carlo simulation is used to generate event samples with single $\tau$ and di-$\tau$ in the final states and originating from $W'$, and $Z'$ decays, respectively, as well as a background process such as $Z\rightarrow\tau\tau$. The parton level events are generated using the \texttt{MadGraph}\_\texttt{aMC}$@$\texttt{NLO 2.9.9} \cite{madgraph} and \texttt{Pythia8}\cite{pythia8} is used for fragmentation and hadronization. The detector simulation and object reconstruction are performed using the fast simulation package Delphes 3.5 \cite{delphes}, with the CMS detector configuration, taking into account the acceptance and expected performance of the detector. 
 
\begin{figure}
 \subfloat{
	\begin{minipage}[c][1\width]{
	   0.5\textwidth}
	   \centering
	   \includegraphics[width=1\textwidth]{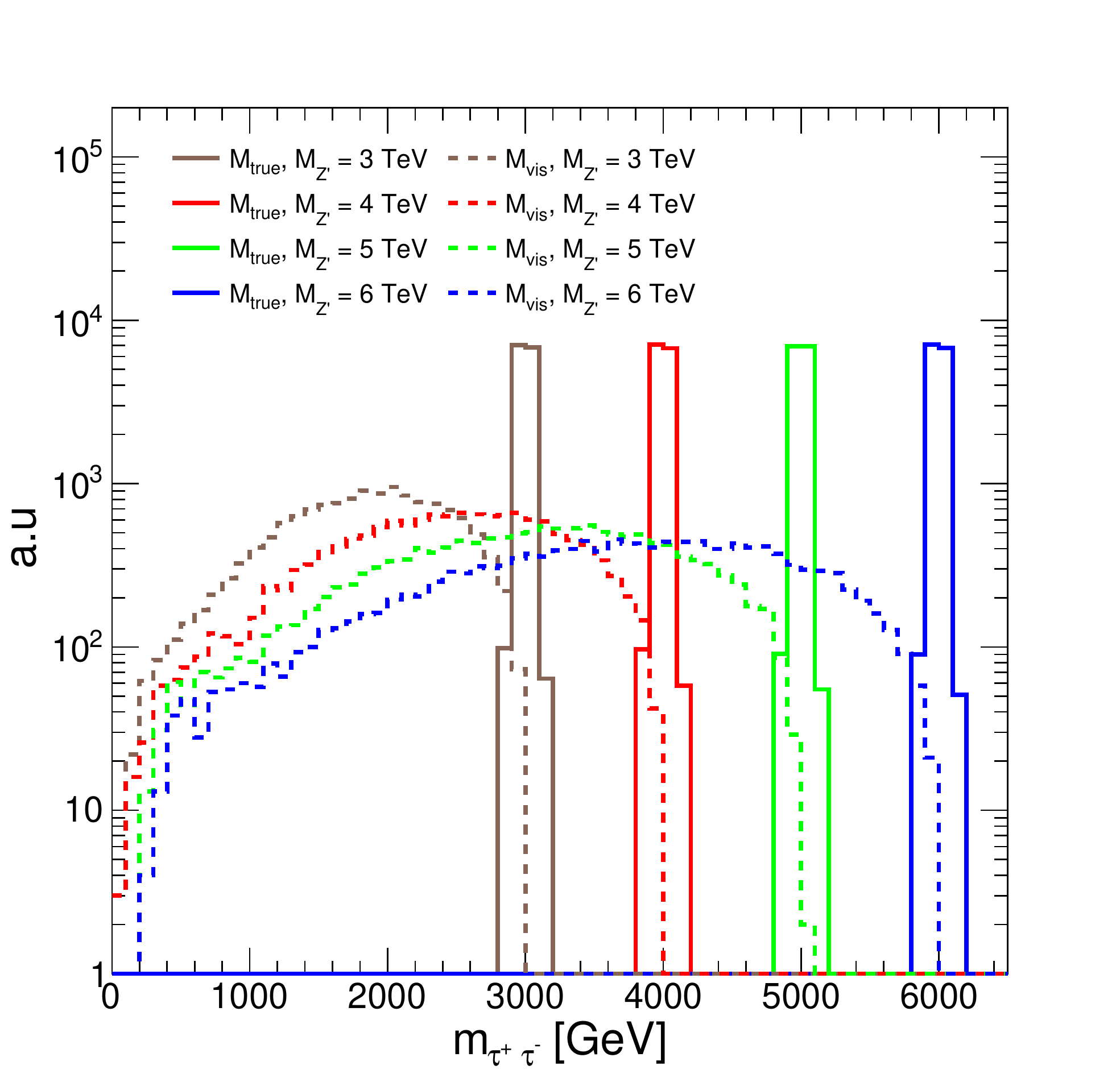}
	\end{minipage}}
 \hfill 	
  \subfloat{
	\begin{minipage}[c][1\width]{
	   0.5\textwidth}
	   \centering
	   \includegraphics[width=1\textwidth]{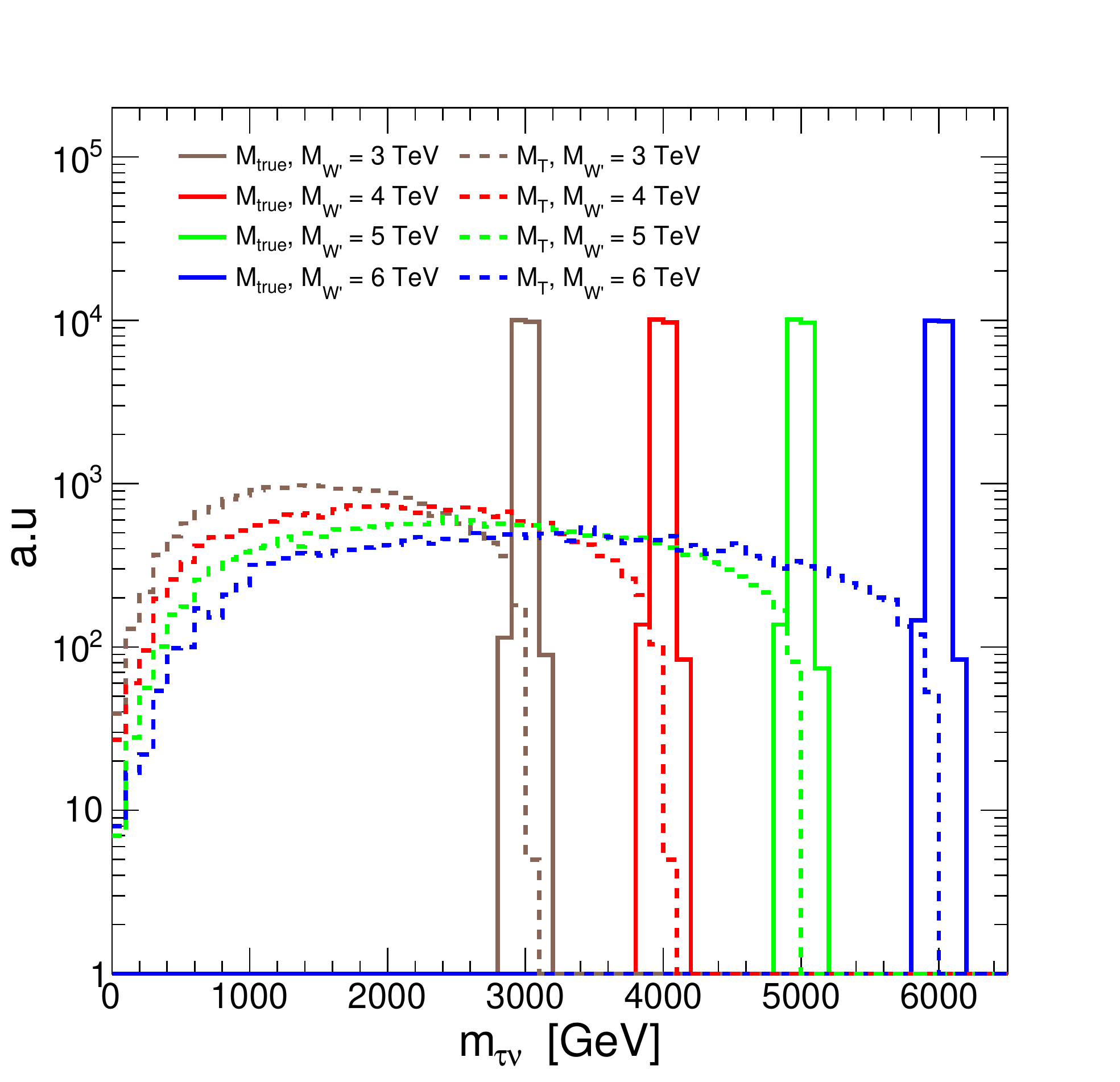}
	\end{minipage}}

\caption{The visible ($M_{vis}$) and generated invariant mass ($M$) of $\tau\tau$ final states (Left). The visible transverse($M_{T}$) and generated mass ($M$) of $\tau\nu_{\tau}$ final states (Right)} 
\label{fig:Fig.1}
\end{figure}

 The heavy gauge bosons are generated by following a simplified model, which extends the SM field content by introducing the massive vector fields $W^{\prime\pm}$ and $Z^{\prime}$\cite{Fuks:2017vtl}. The sensitivity to searches of new heavy bosons is usually explored using a reference model, in which $Z'$($W'$) interacts with the leptons with the same left-handed couplings as the counterpart Z(W) bosons in the standard model. The signal samples are produced for the various resonant masses in the 3-6 TeV range with the 1 TeV interval. The main backgrounds considered for the $Z'\rightarrow\tau\tau$ and $W'\rightarrow\tau\nu_{\tau}$ are the DY($Z/\gamma^* \rightarrow\tau\tau$) and the decay of $W\rightarrow\tau\nu_{\tau}$. The  $p^{miss}_{T} > 120$ GeV criteria is applied to all events to ensure that the neutrino momenta are not pointing in opposite directions. The final states with $\tau$-leptons decaying to hadrons are considered. The sum of the visible components of the tau lepton is required to have the $p^{\tau}_{T} > 80$ GeV and $-2.3 < \eta < 2.3$. The identification efficiencies of the $\tau$-leptons are considered according to \cite{CMS:Tau_id}. Figure \ref{fig:Fig.1} shows the generated and visible invariant mass distributions of the heavy gauge bosons.
In the heavy gauge boson search, the invariant mass can provide a better separation between the signal and the background compared to the visible mass distribution. To reconstruct the invariant mass, we study two kinds of ML algorithms. One is based on supervised learning, and the other exploits an unsupervised learning method. For the di-$\tau$ final states, we study a supervised learning technique, which is a regression method using a deep neural network \cite{Bartschi:2019xlg}. To evaluate the expected improvement in the performance, the ML reconstructed invariant mass is compared to the invariant mass reconstructed using the SVfit algorithm\cite{SVfit}. The single $\tau$ final states have enriched $p^{miss}_{T}$ contribution, where there are two types of neutrino contributions originating from $W'$ and the hadronic $\tau$ decays, respectively. In this case, we study the invariant mass reconstruction using an unsupervised learning method,  by performing regression with a modified adversarial network (mAN). 
\section{Invariant mass of $Z'\rightarrow\tau\tau$}
\label{sec_zmass}
In this section, we study the reconstruction of the invariant mass of $Z'\rightarrow\tau\tau$  using a deep neural network architecture and compare its performance to that of the SVfit algorithm. 
\subsection{DNN regression}
\label{dnn_sec}
High-level keras API gives the architecture for constructing the Deep neural network( DNN). The neural network(NN) used in this study comprises four hidden layers and 107 fully connected neurons. The network architectural details are shown in the Table \ref{table:Table1}. Each layer is re-weighted according to the gradient value of the mean squared loss function. The commonly employed "rectified linear unit (ReLU)" activation function is used. 

\begin{figure}
  \subfloat{
	\begin{minipage}[c][1.25\width]{
	   0.5\textwidth}
	   \centering
	   \includegraphics[width=1\textwidth,height=0.75\textwidth]{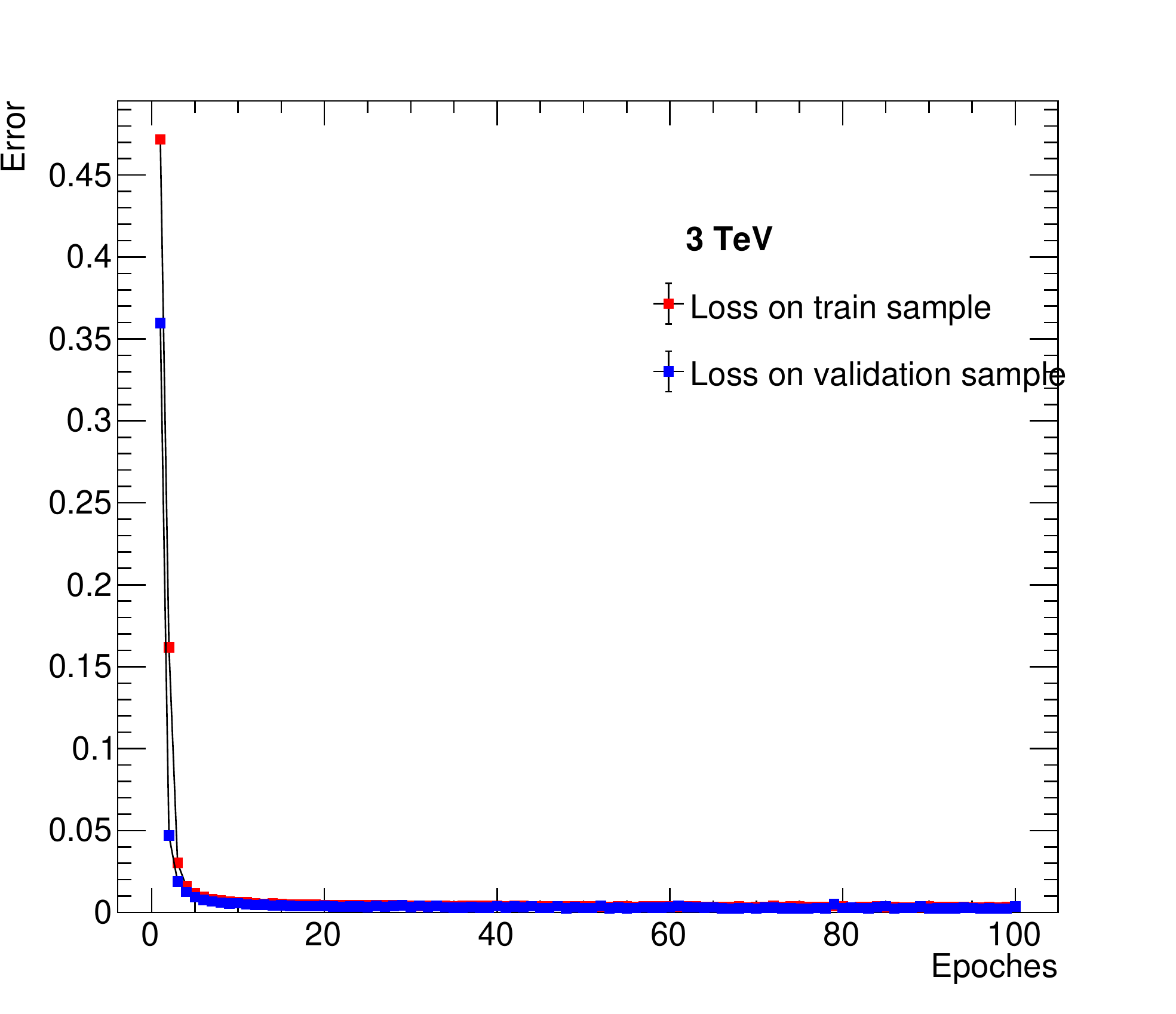}
	\end{minipage}}
  \subfloat{
	\begin{minipage}[c][1.25\width]{
	   0.5\textwidth}
	   \centering
	   \includegraphics[width=1\textwidth,height=0.75\textwidth]{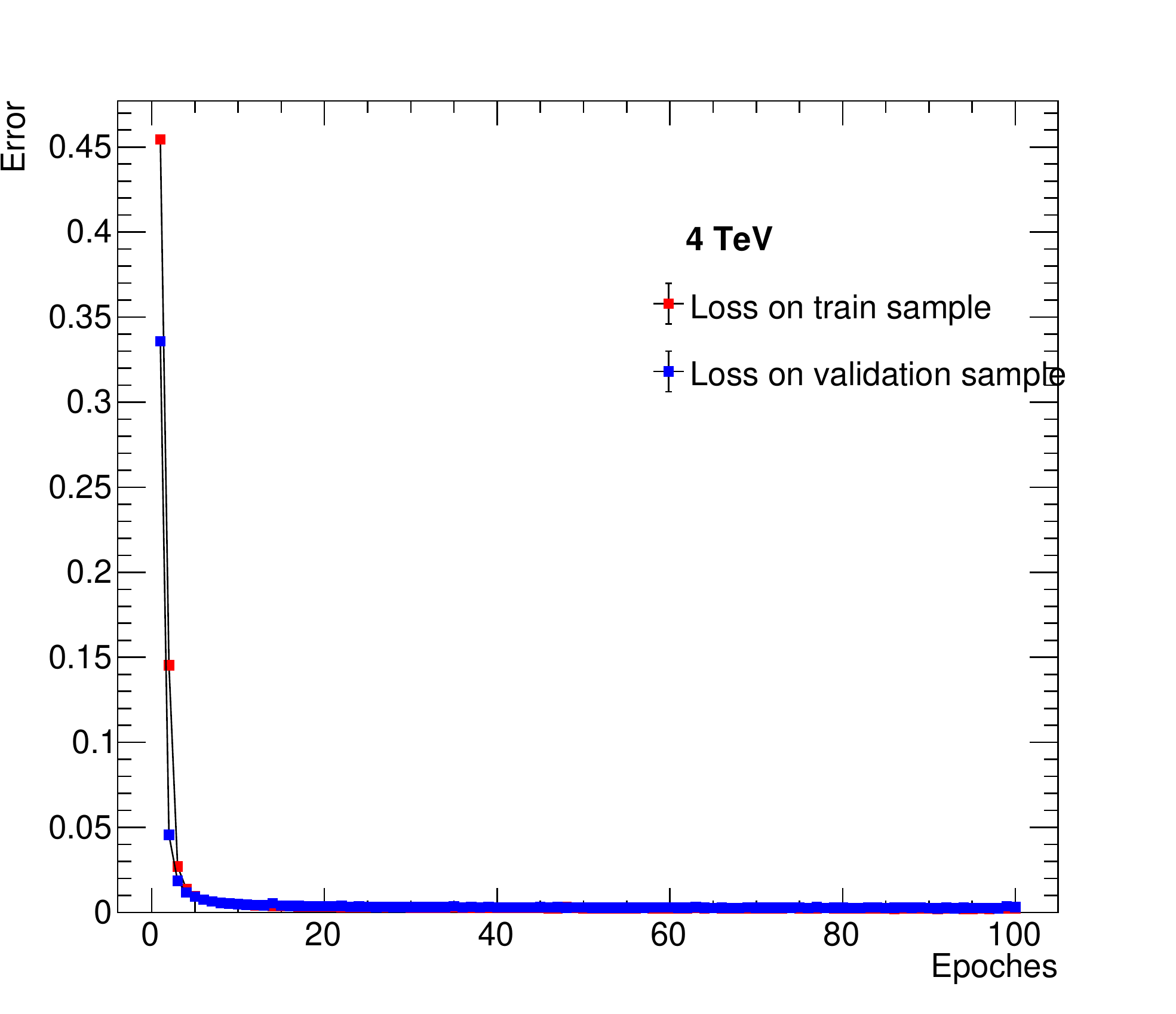}
	\end{minipage}}	
 \vspace{0.1cm}
 \subfloat{
	\begin{minipage}[c][0.15\width]{
	   0.5\textwidth}
	   \centering
	   \includegraphics[width=1\textwidth,height=0.75\textwidth]{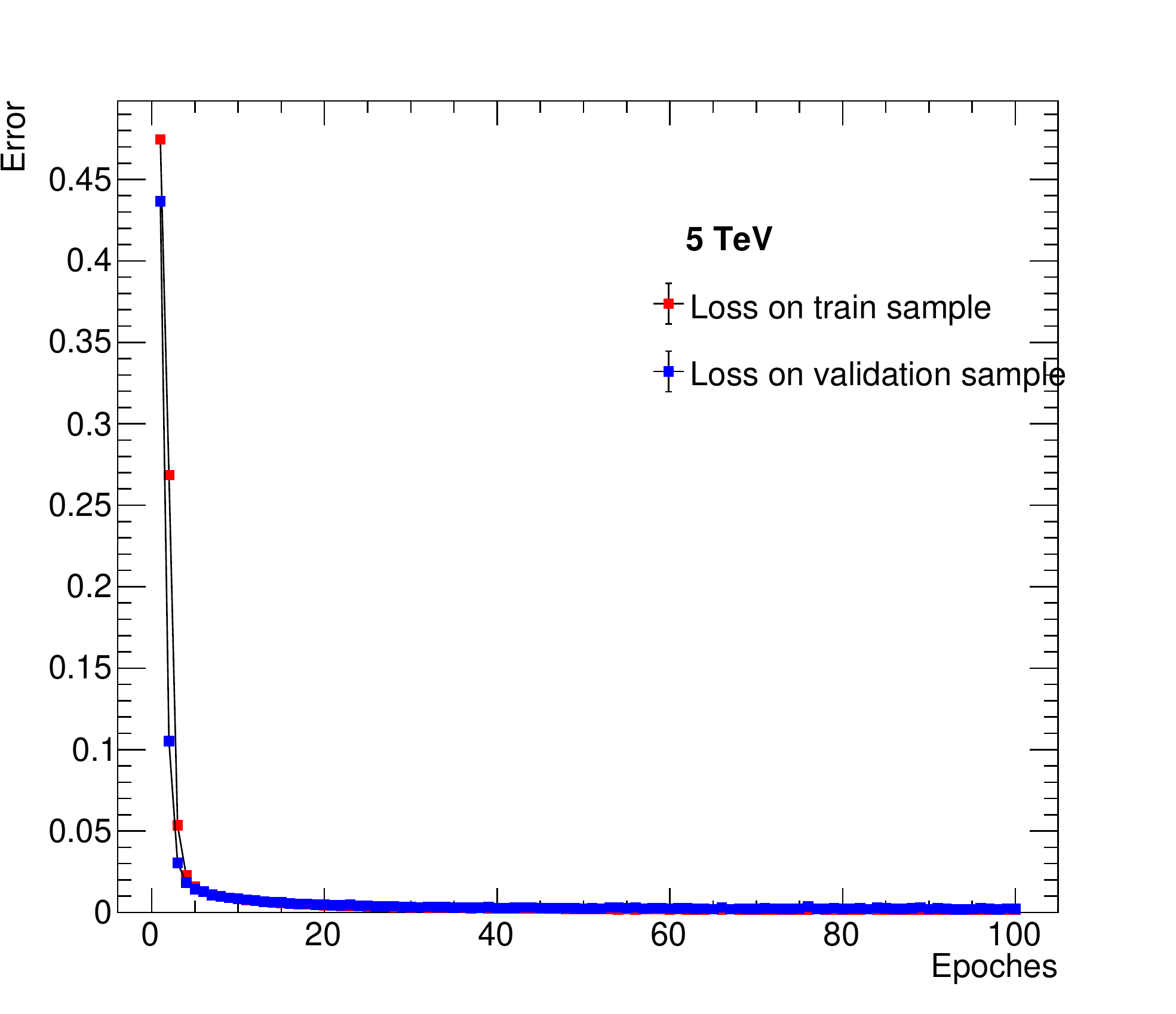}
	\end{minipage}}
\subfloat{
	\begin{minipage}[c][0.15\width]{
	   0.5\textwidth}
	   \centering
	   \includegraphics[width=1\textwidth,height=0.75\textwidth]{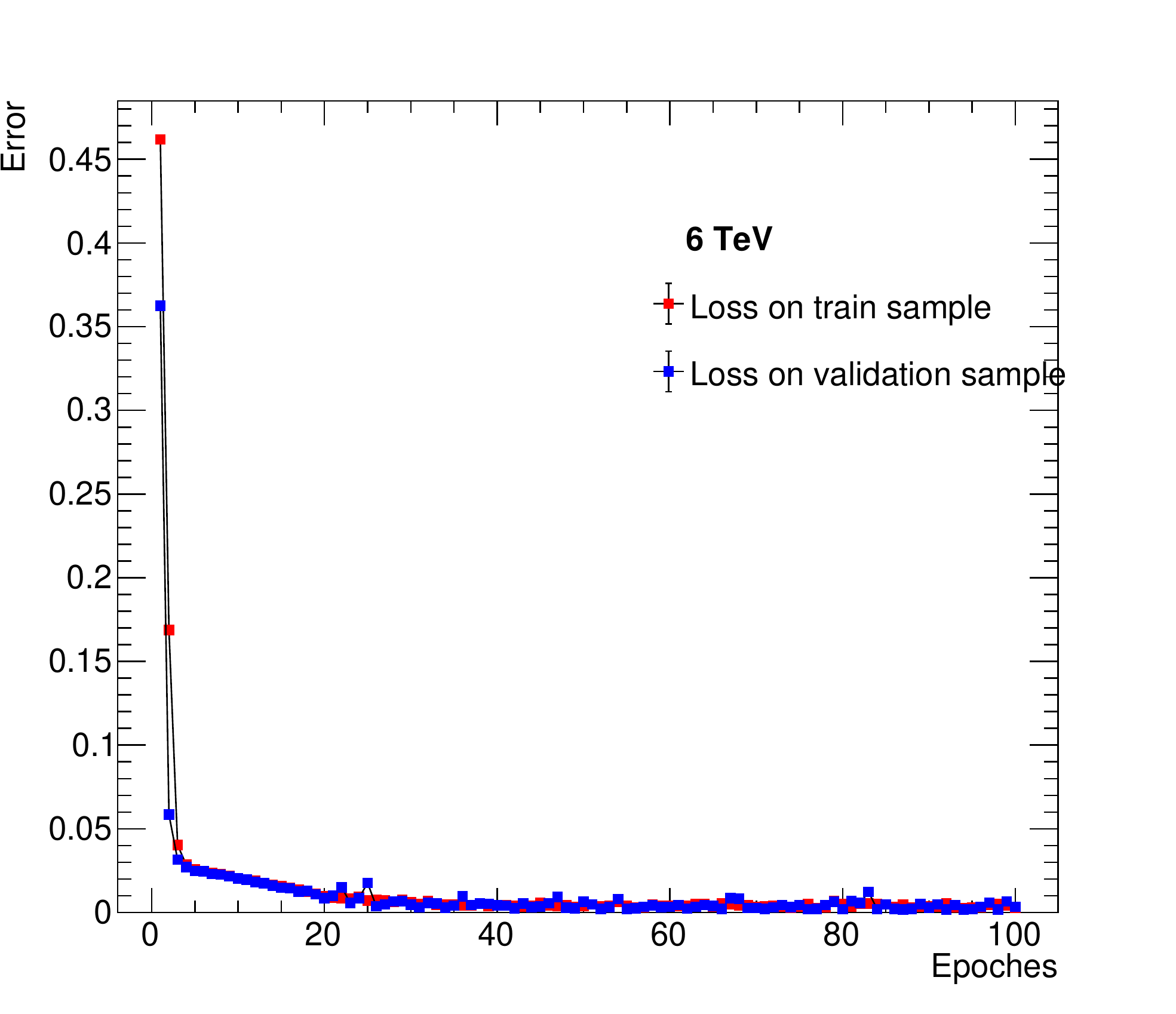}
	\end{minipage}}
 \vspace{2cm}
\caption{The training of DNN for different $Z^{\prime}$ mass values are shown. The validation and training loss values are close for higher epochs, which ensures the training is not over or under-fitted.}
\label{train}
\end{figure}

\begin{table}
\centering
\begin{tabular}{ c c c }
 \textbf{Layer} & \textbf{No. of neurons} & \textbf{Activation Function} \\ 
 Input Layer & 10 & - \\  
 1. Hidden layer & 16  & ReLU \\
 2. Hidden layer & 32 & ReLU \\
 3. Hidden layer & 32 & ReLU \\
 4. Hidden layer & 16 & ReLU \\
 Output layer    & 1 & Linear

\end{tabular}
\caption{The DNN architecture for $Z'\rightarrow\tau\tau$ mass regression}
\label{table:Table1}
\end{table}
The DNN model is trained using ten input variables: Four components of the visible momenta of two taus and the $p_{T}^{miss}$ components. Figure \ref{train} shows the training performance of the DNN regression in terms of the loss over the epoch. The model is trained with the batch size of 128 over an epoch until the loss is saturated. The training samples consist of both background and the signal events. Once the regression model is well trained, the invariant mass distribution is obtained from the test samples, where the test samples consist of only the signal events. 
\subsection{Mass reconstruction using SVfit}
To evaluate the performance of the neural network, the NN reconstructed mass is compared to that of other methods, such as SVfit \cite{SVfit}. SVfit reconstructs the mass of the di-$\tau$ system using a dynamical likelihood technique. The term dynamical likelihood techniques refer to likelihood-based methods used for the reconstruction of kinematic quantities on an event-by-event basis. The inputs to SVfit are the visible decay products of the $\tau$-leptons, $x$, and $y$ components of $p_{T}^{miss}$  as well as its covariance matrix. The $p_{T}^{miss}$ covariance matrix represents the expected resolution of the  $p_{T}^{miss}$ reconstruction in the detector.

\subsection{Results}
\begin{figure}
    \centering
    \includegraphics[width=0.72\textwidth]{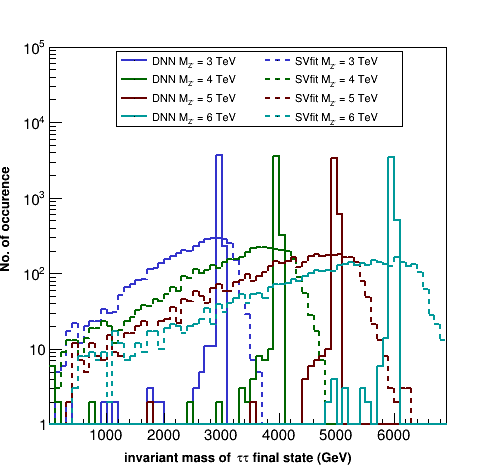}
    \caption{The distribution of $Z'\rightarrow\tau\tau$ invariant mass obtained from the DNN regression and SVfit method. }
    \label{resDNN}
\end{figure}

The reconstructed invariant mass for different $m_{Z'}$ are shown in Fig. \ref{resDNN}, and the resolution of the reconstructed invariant mass distributions obtained from the SVfit and DNN algorithm are compared in Table \ref{resTable}. It is observed that the resolution improves significantly by DNN regression, where the resolution is defined as the ratio of root mean square deviation (GeV) and the mean(GeV). In addition, it is to be noted that the reconstruction using DNN regression is faster than the SVfit algorithm. 
\begin{table}[ht]
\begin{tabular}{c|c|c} 
  \hline
  Mass (TeV) & resolution from DNN (r.m.s/$\mu$) & resolution from SVfit (r.m.s/$\mu$) \\ 
  \hline
  3  & 0.0291 & 0.290 \\ 
  \hline
  4  & 0.0145 & 0.275 \\ 
  \hline
  5  & 0.0238  & 0.274 \\ 
  \hline
  6  & 0.0130 & 0.268 \\ 
  \hline
\end{tabular}
\caption{The comparison of resolutions for the invariant mass distributions obtained from DNN and SVfit methods.}
\label{resTable}
\end{table}


\section{Invariant mass of $W'\rightarrow\tau\nu_{\tau}$}
\label{sec_wmass}
Unlike the $Z'\rightarrow\tau\tau$ process where we have ten variables, such as the 4-momenta of each tau and the components of missing transverse energy to fit the DNN regression model, there are only six variables available for single tau final state. The unsupervised learning models have good pattern recognition skills. The adversarial neural network is one of them, which can learn the invariant mass distribution from the given data and help to reproduce the same distribution. 
\subsection{DNN regression}
To compare the results of the adversarial networks, we perform regression using a deep neural network as discussed in the section \ref{dnn_sec}. The best-optimized model for $W'$ is shown in Table \ref{table:Table2}.
\begin{table}
\centering

\begin{tabular}{ c c c }
 \textbf{Layer} & \textbf{No. of nurons} & \textbf{Activation Function} \\ 
 Input Layer & 6 & - \\  
 1. Hidden layer & 32  & ReLU \\
 2. Hidden layer & 16 & ReLU \\
 3. Hidden layer & 16 & ReLU \\
 4. Hidden layer & 32 & ReLU \\
 Output layer    & 1 & Linear

\end{tabular}
\caption{DNN architecture for $W'\rightarrow\tau\nu_{\tau}$ mass regression}
\label{table:Table2}
\end{table}

\begin{figure}
  \subfloat{
	\begin{minipage}[c][1\width]{
	   0.5\textwidth}
	   \centering
	   \includegraphics[width=1\textwidth]{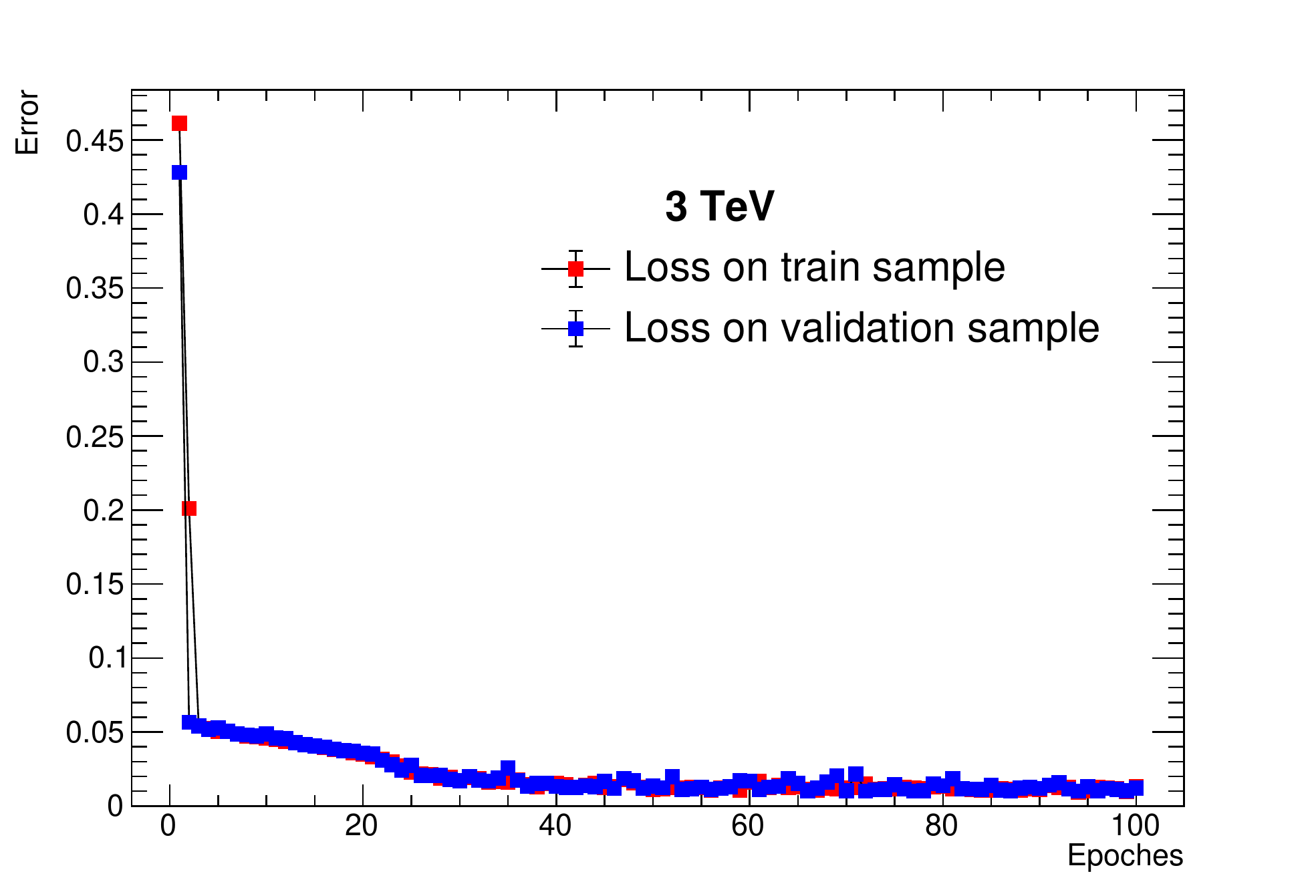}
	\end{minipage}}
  \subfloat{
	\begin{minipage}[c][1\width]{
	   0.5\textwidth}
	   \centering
	   \includegraphics[width=1\textwidth]{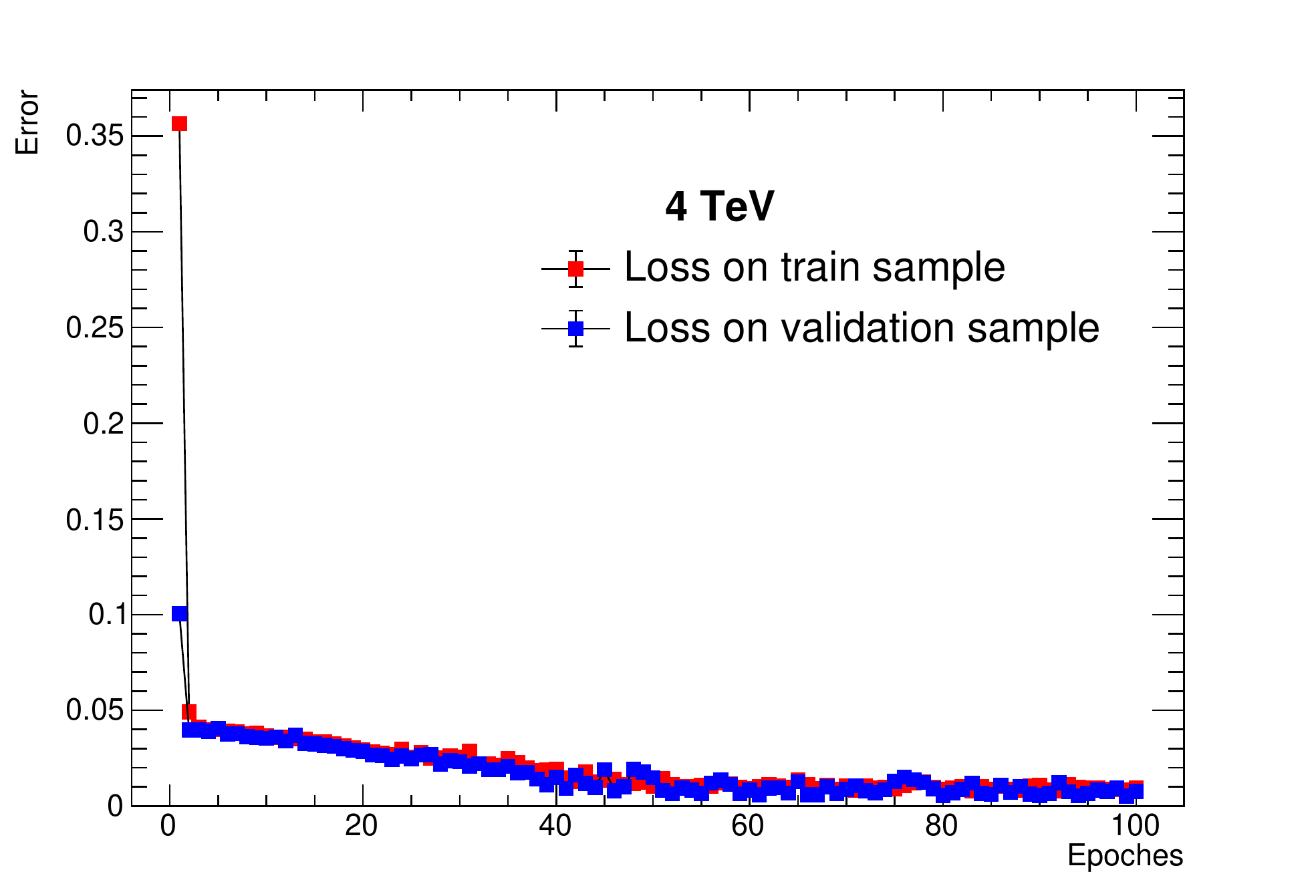}
	\end{minipage}}
 \vspace{0.1mm}
 \subfloat{
	\begin{minipage}[c][0.25\width]{
	   0.5\textwidth}
	   \centering
	   \includegraphics[width=1\textwidth]{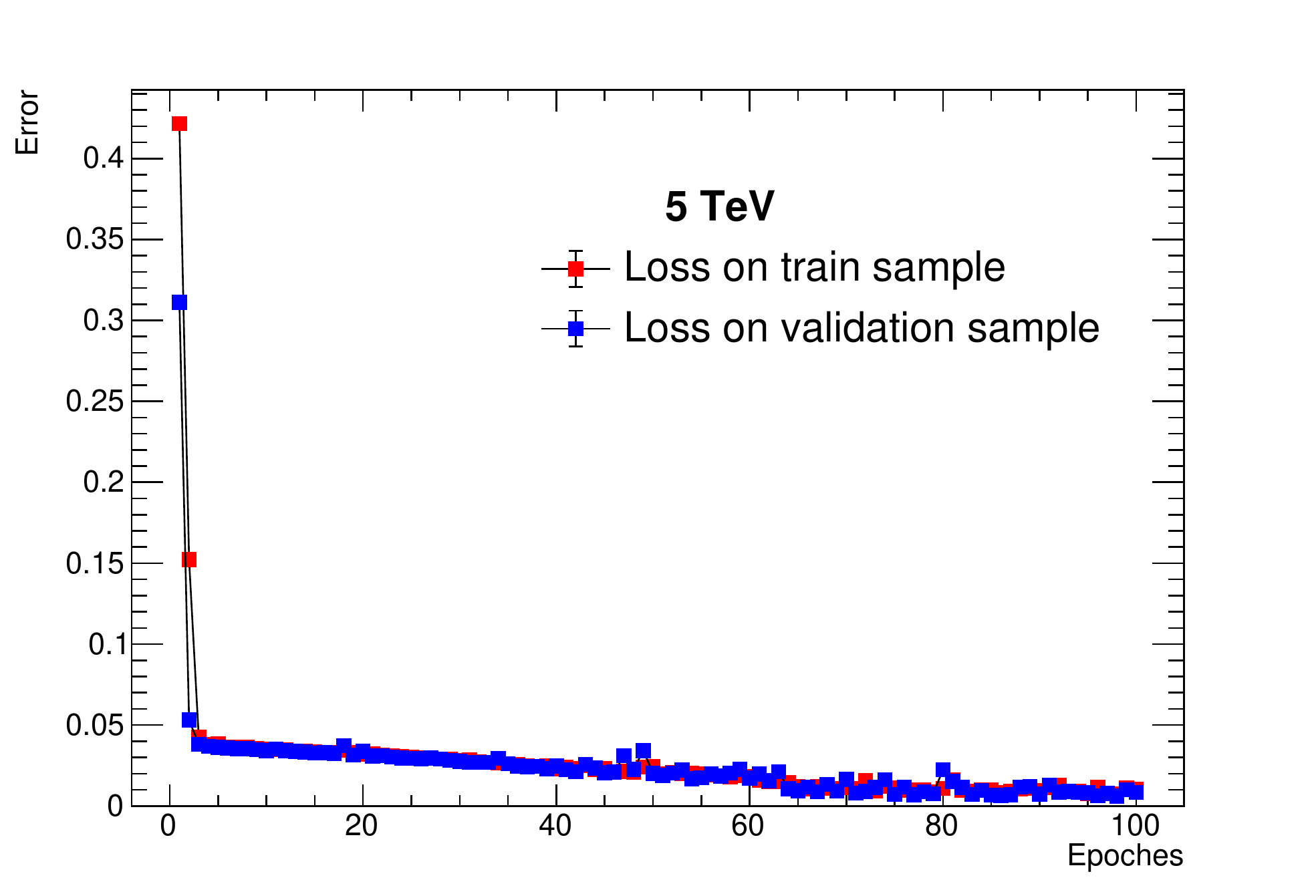}
	\end{minipage}}
\subfloat{
	\begin{minipage}[c][0.25\width]{
	   0.5\textwidth}
	   \centering
	   \includegraphics[width=1\textwidth]{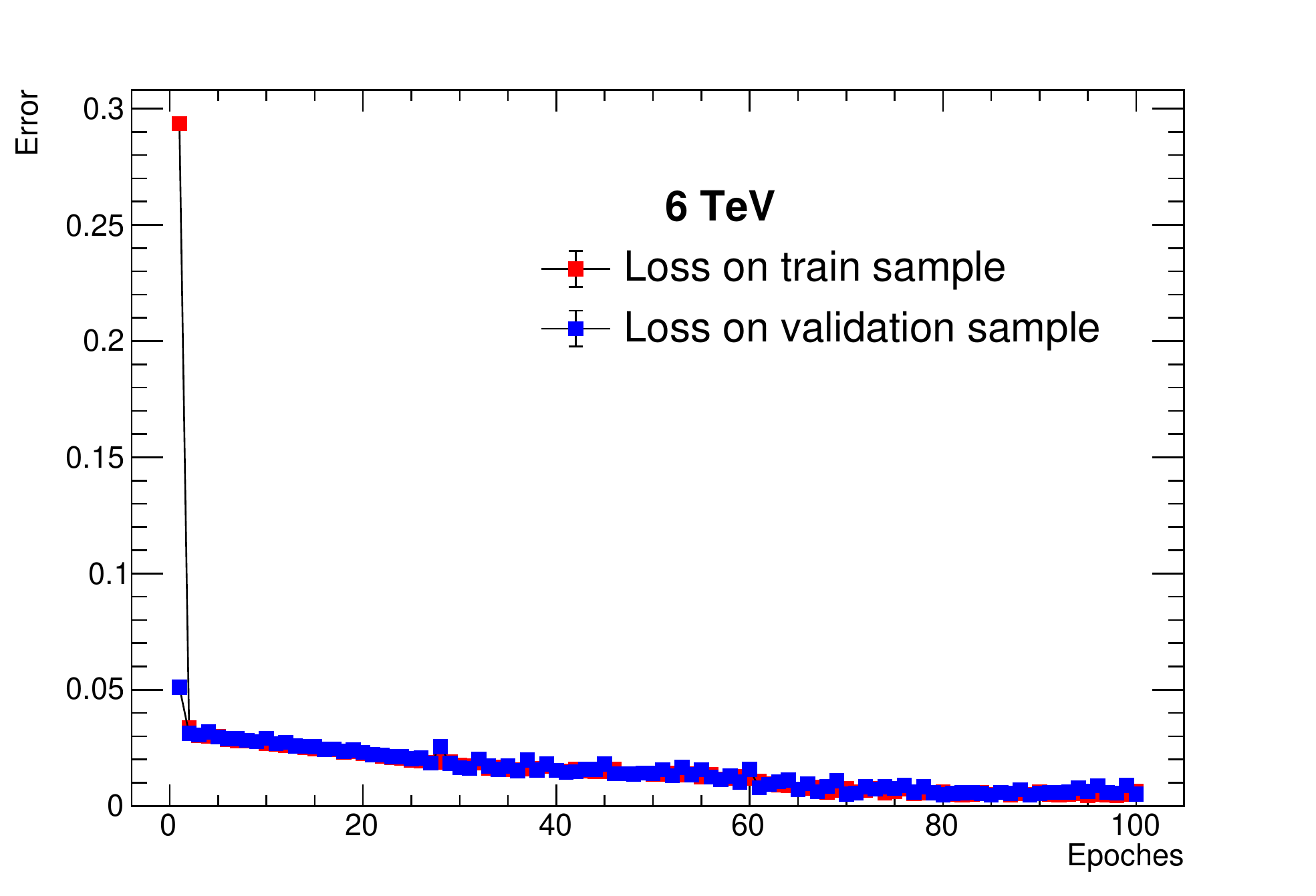}
	\end{minipage}}
 \vspace{1cm}
\caption{The training of DNN for different $W^{\prime}$ mass are shown. The validation and training loss values are close for higher epochs, which ensures the training is not over or under-fitted.}
\label{trainw}
\end{figure}
The DNN model is trained using six input variables: Four components of the visible momenta of the tau and the $p_{T}^{miss}$ components. Figure \ref{trainw} shows the training performance of the DNN regression in terms of the loss over the epoch. After 100 epochs, the loss is saturated by obtaining the optimized  NN model. Once the DNN regression model is trained, the invariant mass distribution is obtained from the test samples, where the samples only consist of signal events. 
\subsection{Adversarial network regression}
\begin{figure}
    \centering
    \includegraphics[width=1\textwidth]{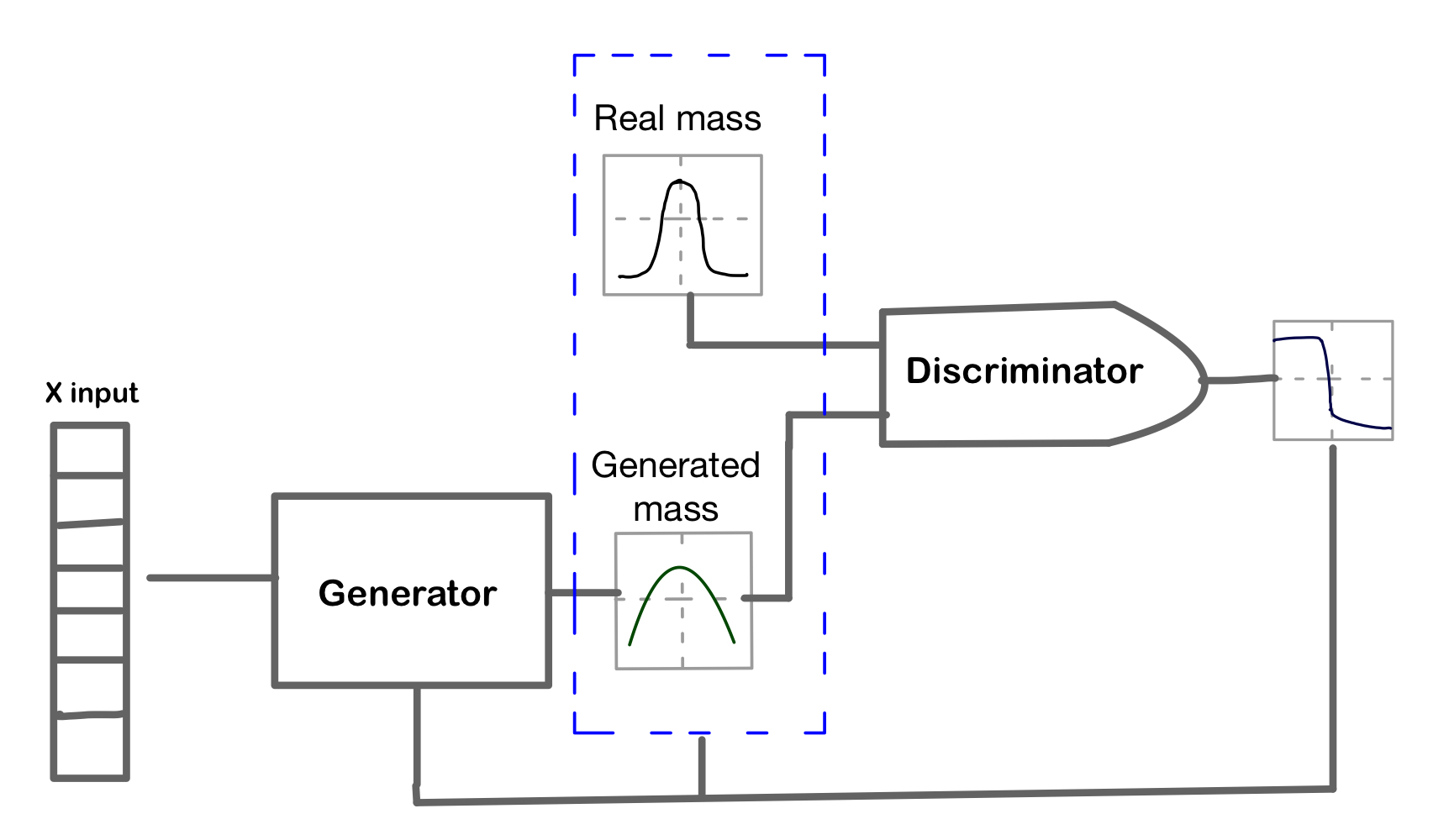}
    \caption{A schematic diagram of the adversarial neural network used for performing mass regression of $W'\rightarrow\tau\nu$}
    \label{gan}
\end{figure}

\begin{figure}
\centering
\includegraphics[width=1\textwidth]{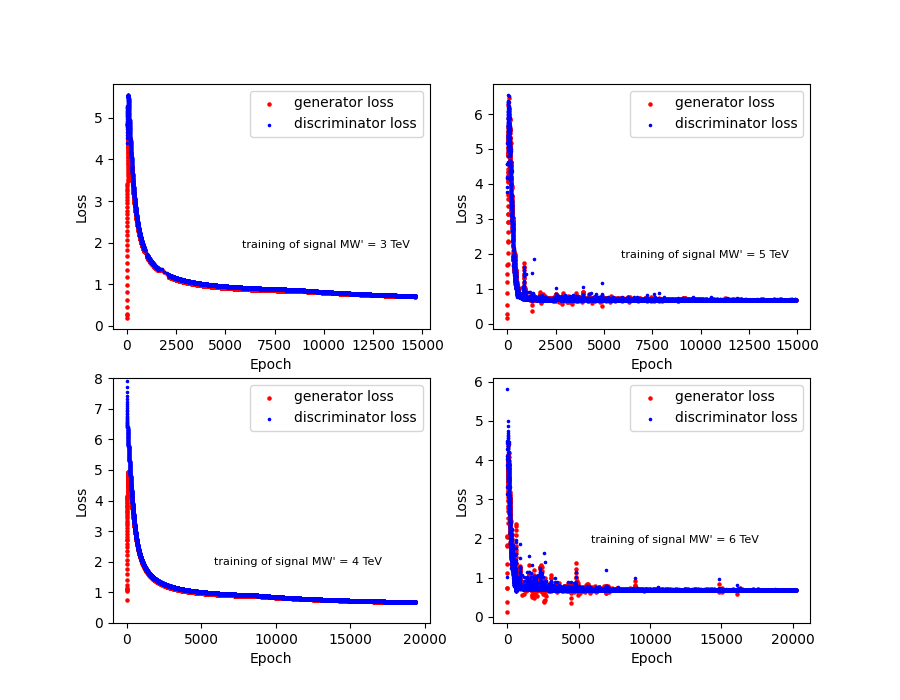}
\captionof{figure}{The training performances  of mAN for different $W^{\prime}$ mass are shown. The plots show the  loss values for discriminator and generator. }
\label{sep}
\end{figure}

The adversarial network consists of two neural network components, generator(G) and discriminator(D), as shown in Fig \ref{gan}. A generator neural network component, which is the same as the DNN model used in the previous section, has the input layer with the components of the visible tau momentum and $p_{T}^{miss}$ as input variables and one output node. The discriminator is a classification network, having the input as the simulated mass of $W'$ and the output value of the generator network. 
The architectures of these two networks are described in Table \ref{tab:ganarc}.
\begin{table}[]
    \centering
    \begin{tabular}{c|c|c|c}
    Generator & & & \\
         & \textbf{Layers} & \textbf{No. of Nurons} & \textbf{Activation Function} \\
         & Input layer & 6 & ReLU \\
         & hidden layer1 & 32 & ReLU \\
         & hidden layer2 & 16 & ReLU\\
         & hidden layer3 & 16 & ReLU\\
         & hidden layer3 & 32 & ReLU\\
         & output layer & 1 & linear\\
    Discriminator & & & \\
        & Input layer & 1 & ReLU\\
        & hidden layer1 & 256 & Leaky ReLU($\alpha$ = 0.5)\\
         & hidden layer2 & 128 & Leaky ReLU($\alpha$ = 0.5)\\
         & output layer & 1 & sigmoid\\
    \end{tabular}
    \caption{The detailed architecture  of the modified Adversarial Network model.}
    \label{tab:ganarc}
\end{table}
Unlike the DNN regression, the generator learns from the output value of the discriminator, because of which it is considered as an unsupervised learning \cite{GANref}. However, the output of the generator and the true simulated mass of the $W'$ boson are fed to the discriminator, which learns to discriminate the true simulated mass of the $W'$ boson from that of the mass obtained by the generator. After a certain iteration of training,  both the generator and the discriminator find Nash's equilibrium when the output of the generator matches that of the true simulated value and the discriminator output is approximately 0.5. The discriminator has the input of simulated mass distribution ($\mu_M$) and the generated mass distribution from generator network ($\mu_G$). The value $D(x)$ is the discriminator's estimate of the probability that real data ($\mu_M$) instance $x$ is real and $G(x_{input})$ is the generator's output with the training feature ($x_{input}$). The loss function of the adversarial network is defined similar to the generative adversarial network (GAN)\cite{Ganref}, except we modify the network by adding mean square error loss function as shown in Eq.\ref{eq1}. This helps to train the generator with the input variables.

\begin{equation}
L(D,G) = \mathop{\mathbb{E}}_{x\sim \mu_M}\log{D(x)} + \mathop{\mathbb{E}}_{x_{input}\sim \mu_G}\log\left(1-D(G(x_{input}))\right) + \frac{(G(x_{input}) - x)^2}{2}
\label{eq1}
\end{equation}
The training performance of adversarial neural network is obtained from the generator and discriminator loss functions. Figure \ref{sep} shows the values of the loss over iterations for both generator and discriminator, which are saturated and coinciding with each other after sufficient number of iterations. This ensures that the model is fitted properly.   Once the mAN model is trained well, the invariant mass distribution obtained from the test samples, where the samples only consists of the signal events.

\subsection{Results}
\begin{figure}
    \centering
    \includegraphics[width=0.72\textwidth]{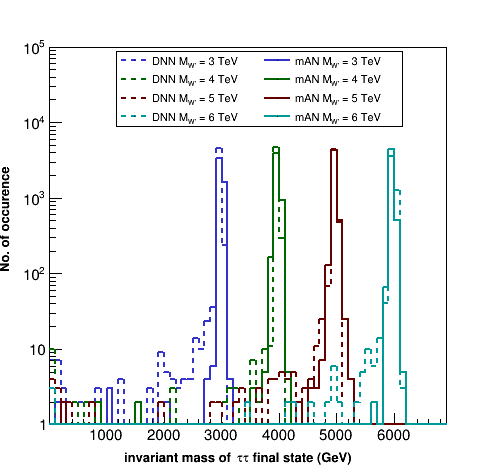}
    \caption{The invariant mass distributions  of $W'$ obtained from the DNN regression and modified adversarial network. }
    \label{resGAN}
\end{figure}

Figure \ref{resGAN} shows the invariant mass distributions  obtained by the adversarial model and is compared to that obtained with the DNN regression model. The comparison of the resolutions of the invariant mass distributions, defined as $\text{r.m.s}/\mu$, obtained from the adversarial and DNN algorithm, are presented in Table \ref{gresTable}. The comparisons of resolutions are presented for different $m_{W'}$ values. From the Fig. \ref{resGAN} and Table \ref{gresTable}, it is observed that  the mass regression using the  mAN model not only reconstructs the full invariant mass of $W'\rightarrow\tau\nu$ system, where a part of the energy is missing due to neutrinos, but also improves its resolution significantly in comparison to that of the regression using a DNN model.  

\begin{table}
\centering
\begin{tabular}{c|c|c} 
  \hline
  Mass (TeV) & DNN (rms/$\mu$) & adversarial(rms/$\mu$)\\ 
  \hline
  3   & 0.072 & 0.0452\\ 
  \hline
  4   & 0.066 & 0.0368 \\ 
  \hline
  5   & 0.057 & 0.0395 \\ 
  \hline
  6   & 0.049 & 0.0163 \\ 
  \hline
\end{tabular}
\caption{The comparison of resolutions of the invariant mass distributions obtained from DNN and adversarial neural network}
\label{gresTable}
\end{table}

\section{Summary}
We studied different ML-based algorithms to reconstruct invariant mass of high mass resonances decaying to $\tau$-lepton final states. It is found that a DNN-based mass regression provides better performance, in terms of mass resolution, in reconstructing the invariant mass of the $Z'\rightarrow\tau\tau$ system in comparison to traditional likelihood-based algorithms, such as SVfit. It is also relatively faster in terms of computing time. We also studied an unsupervised method, employing a modified adversarial network model, to reconstruct the full invariant mass of the $W'\rightarrow\tau\nu$ decays, where the $\tau$ decays to hadrons and neutrino. The performance of the adversarial model is compared to that of the mass regression obtained using a DNN model with the same set of input variables. The invariant mass obtained with the adversarial model not only restores the mass peak but also significantly improves its resolution  in comparison to that of the DNN model. We expect that such a reconstruction of the invariant mass will provide well-separated distributions against the standard model backgrounds and will significantly improve the search capacity of the $W'\rightarrow\tau\nu$ processes. Furthermore, this technique can also be applicable for reconstructing the invariant mass of $W'\rightarrow e\nu$, $\mu\nu$ processes as well as that of the charged Higgs boson ($H^+\rightarrow\tau\nu$), significantly enhancing their search capabilities. \\
\newpage

\printbibliography

\end{document}